\documentclass[prl,twocolumn,nofootinbib,floatfix,subeqn]{revtex4-1}
\usepackage[usenames]{color}
\usepackage{amsmath}
\usepackage{amssymb}
\usepackage{float}
\usepackage{graphicx}
\usepackage{xspace}
\usepackage{hyperref}
\usepackage{hhline}
\usepackage{bbold}

\newcommand{\change}[1]{{\color{black}#1\color{black}}}

\newcommand{\mycite}[1]{\cite{#1}}

\long\def\symbolfootnote[#1]#2{\begingroup%
\def\thefootnote{\fnsymbol{footnote}}\footnotetext[#1]{#2}\endgroup}

\begin{document}

\title{The `Higgs' Amplitude Mode at the Two-Dimensional Superfluid-Mott Insulator Transition}


\author{Manuel~Endres$^{1,*}$}%
\author{Takeshi~Fukuhara$^{1}$}%
\author{David~Pekker$^{2}$}%
\author{Marc~Cheneau$^{1}$}%
\author{Peter~Schau\ss$^{1}$}%
\author{Christian~Gross$^{1}$}%
\author{Eugene Demler$^{3}$}%
\author{Stefan~Kuhr$^{4}$}%
\author{Immanuel~Bloch$^{1,5}$}%

\date{23 April 2012}

\affiliation{\vspace{0.2cm}$^1$Max-Planck-Institut f\"{u}r Quantenoptik, 85748 Garching, Germany}
\affiliation{$^2$Department of Physics, Caltech University, Pasadena, California 91125, USA}
\affiliation{$^3$Physics Department, Harvard University, Cambridge, Massachusetts 02138, USA}
\affiliation{$^4$University of Strathclyde, SUPA, Glasgow G4 0NG, United Kingdom}%
\affiliation{$^5$Ludwig-Maximilians-Universit\"{a}t, 80799 M\"{u}nchen, Germany}%

\begin{abstract}
Spontaneous symmetry breaking plays a key role in our understanding of nature. In a relativistic field theory, a broken continuous symmetry leads to the emergence of two types of fundamental excitations: massless Nambu-Goldstone modes and a massive `Higgs' amplitude mode.  An excitation of Higgs type is of crucial importance in the standard model of elementary particles \mycite{Weinberg:1996} and also appears as a fundamental collective mode in quantum many-body systems \mycite{Sachdev:2011}.  Whether such a mode exists in low-dimensional systems as a resonance-like feature or becomes over-damped through coupling to Nambu-Goldstone modes has been a subject of theoretical debate \mycite{Chubukov:1994,Sachdev:1999, Zwerger:2004, Lindner:2010, Sachdev:2011, Podolsky:2011}. Here we \change{experimentally} reveal and study a Higgs mode in a two-dimensional neutral superfluid close to the transition to a Mott insulating phase. We unambiguously identify the mode by observing \change{the expected softening of the onset of spectral response when approaching the quantum critical point}. In this regime, our system is described by an effective relativistic field theory with a two-component quantum-field \mycite{Altman:2002, Sachdev:2011, Polkovnikov:2005}, constituting a minimal model for spontaneous breaking of a continuous symmetry. Additionally, all microscopic parameters of our system are known from first principles and the resolution of our measurement allows us to detect excited states of the many-body system at the level of individual quasiparticles. This allows for an in-depth study of Higgs excitations, which also addresses the consequences of reduced dimensionality and confinement of the system. Our work constitutes a first step in exploring emergent relativistic models with ultracold atomic gases.
\end{abstract}

\maketitle

\symbolfootnote[1]{Electronic address: {\bf manuel.endres@mpq.mpg.de}}


Higgs modes are amplitude oscillations of a quantum field and appear as collective excitations in quantum many-body systems as a consequence of spontaneous breaking of a continuous symmetry. Close to a quantum critical point, the low-energy physics of such systems is in many cases captured by an effective Lorentz invariant critical theory \mycite{Sachdev:2011}.  The minimal version of such a theory describes the dynamics of a complex order parameter $\Psi=|\Psi|e^{i\phi}$ near a quantum phase transition between an ordered ($|\Psi|>0$) and a disordered phase ($|\Psi|=0$). Within the ordered phase, the classical energy density has the shape of a Mexican hat (Fig.\,1a) and the order parameter takes on a non-zero value in the minimum of this potential. Hereby, its phase $\phi$ acquires a definite value through spontaneous breaking of the rotation symmetry (i.e., $U(1)$ symmetry). Expanding the field around the symmetry broken ground state leads to two types of modes:  a Nambu-Goldstone mode and a Higgs mode related to phase and amplitude variations of $\Psi$, respectively (Fig.\,1a). In contrast to the phase mode, the amplitude mode has a finite excitation gap (i.e., a finite mass), which is expected to show a characteristic softening when approaching the disordered phase (Fig.\,1a).  The sketched minimal model of an order parameter with $N=2$ components belongs to a class of $O(N)$ relativistic field theories, which are essential for the study of quantum phase transitions \mycite{Sachdev:2011}

 \begin{figure}
    \centering
   \includegraphics[width=0.8\columnwidth]{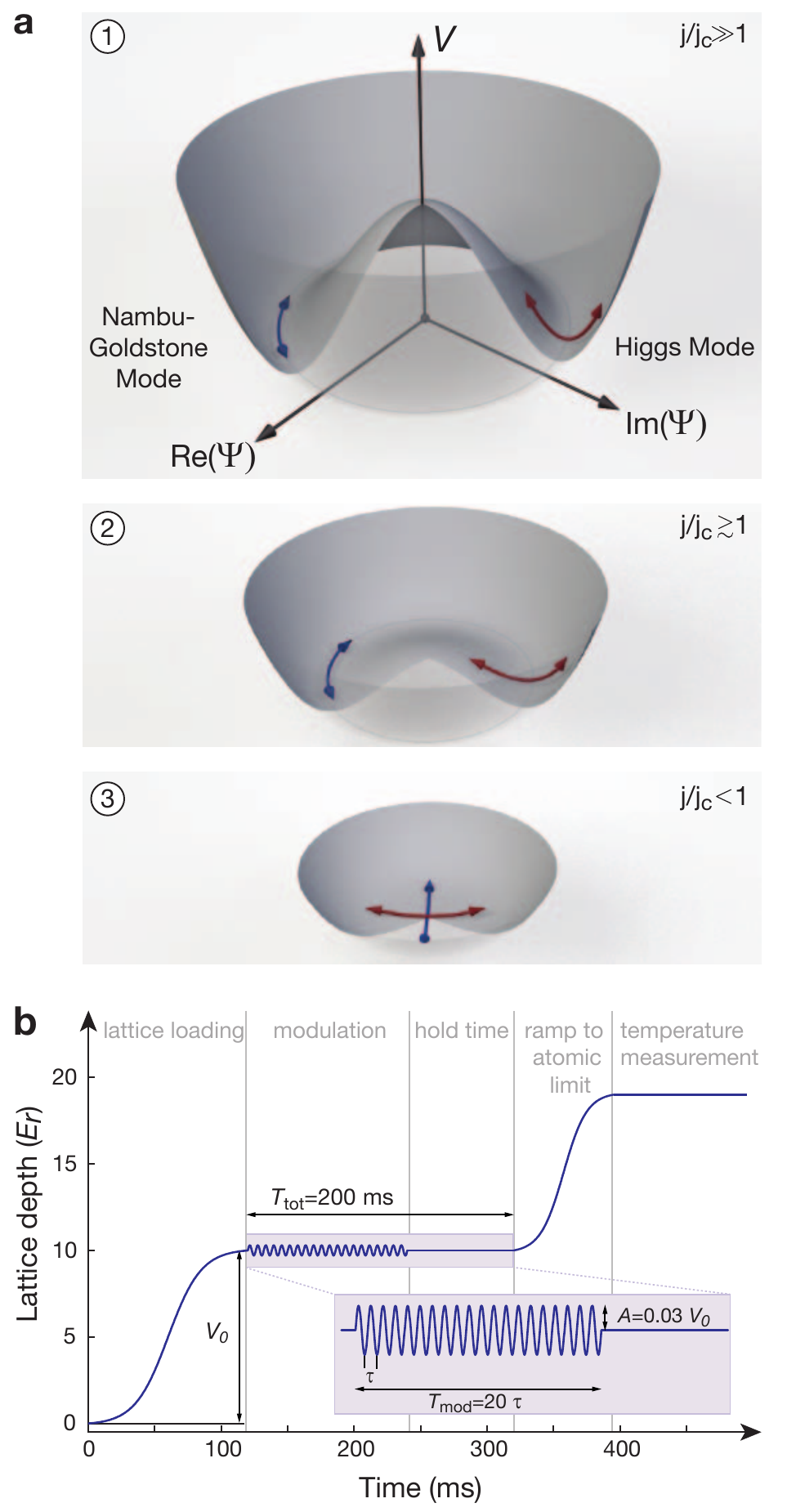}
    \caption{{\bf Illustration of the Higgs mode and experimental sequence.} \textbf{a}, Classical energy density $V$ as a function of the order parameter $\Psi$. Within the ordered (superfluid) phase, Nambu-Goldstone and Higgs modes arise from phase and amplitude modulations (blue and red arrows in panel 1). With the coupling $j=J/U$ (see main text) approaching the critical value $j_c$, the energy density transforms into a function with a minimum at $\Psi=0$ (panels 2-3). Simultaneously, the curvature in the radial direction softens, leading to a characteristic reduction of the excitation frequency for the Higgs mode (reduced bending of the red arrow in panel 2). In the disordered (Mott-insulating) phase, two gapped modes exist, corresponding to particle and hole excitations in our case (red and blue arrow in panel 3). \textbf{b}, The Higgs mode can be excited with a periodic modulation of the coupling $j$, which amounts to a shaking of the classical energy density potential. In the experimental sequence, this is realized by a modulation of the optical lattice potential (see main text for details).}
  \end{figure}
Despite the fundamental nature of the amplitude mode, a full theoretical understanding of it has not yet been achieved.
In particular, the decay of the amplitude mode into lower lying phase modes, especially in two dimensions, has led to a 
considerable theoretical interest concerning the observability of the mode: does a resonance-like feature of the amplitude mode persist, or does the decay result in a low-frequency divergence \mycite{Chubukov:1994, Sachdev:1999, Zwerger:2004, Sachdev:2011, Lindner:2010, Podolsky:2011, Pollet:2012}?

The earliest experimental evidence for a Higgs mode stems from Raman scattering in a superconducting charge-density wave compound showing an unexpected peak \mycite{Sooryakumar:1980}, which was later interpreted as a signal of an amplitude mode \mycite{Littlewood:1981}. Further examples of experiments in solid-state systems can be found in Ref.\,\cite{Podolsky:2011}. Importantly, none of these experiments have studied the mode spectrum across a quantum phase transition, except for neutron scattering experiments on quantum antiferromagnets \mycite{Rueegg:2008}. In contrast to the work presented here, a resonance-like response of an amplitude mode is undoubtedly expected in these systems, because the phase transition occurs in three dimensions.

Ultracold bosonic atoms in optical lattices offer unique possibilities to study quantum phase transitions in a reduced dimensionality \mycite{Bloch:2008}. These systems are nearly ideal realizations of the Bose-Hubbard model, which is parametrized by a tunnelling amplitude $J$ and an on-site interaction energy $U$ (see Methods). The coupling parameter $j=J/U$ is easily tunable via the lattice depth and the dimensionality of the system can be reduced by suppressing hopping in a certain direction \mycite{Spielman:2007, Gemelke:2009, Bakr:2010,Sherson:2010}. At a critical coupling $j_c$ and commensurate filling, the system undergoes a quantum phase transition from a superfluid (ordered) to a Mott insulating (disordered) phase \mycite{Bloch:2008}, which is described by an $O(2)$ relativistic field theory \mycite{Altman:2002, Polkovnikov:2005,Sachdev:2011}. A number of theoretical works have studied the Higgs mode in this system \mycite{Altman:2002, Polkovnikov:2002, Sengupta:2005, Huber:2007, Huber:2008, Menotti:2008, Grass:2011, Pollet:2012}. In particular, it has been argued that a modulation of the lattice depth can reveal a Higgs mode even in a two-dimensional system \mycite{Podolsky:2011,Pollet:2012}.

Previous experiments using a lattice modulation amplitude of $20\%$  were unable to identify the gapped amplitude mode \mycite{Stoeferle:2004, Schori:2004}, most likely owing to the strong non-linear drive \mycite{Kollath:2006}. A recent theoretical analysis of experiments using Bragg scattering in three-dimensional superfluids interpreted parts of the measured spectrum to be the result of non-linear coupling to a short-wavelength amplitude mode \mycite{Bissbort:2011}. Here we experimentally study the long-wavelength and low-energy response, which is described by a relativistic field theory at the quantum critical point.

Our experiment began with the preparation of a two-dimensional degenerate gas of $^{87}$Rb atoms in a single anti-node of an optical standing wave \mycite{Endres:2011}. To realize different couplings $j$, we loaded the two-dimensional gas into a square optical lattice with variable depth $V_0$ (Fig.\,1b). With our trapping parameters and atom numbers (see Methods), the density in the center of the trap is typically one atom per lattice site. We then modulated the lattice depth with an amplitude of $3\%$ at variable frequencies $\nu_{\rm{mod}}$. The modulation time was set to 20 oscillations cycle, thus avoiding an unwanted enhanced response at higher frequencies present in experiments with fixed modulation time \mycite{Stoeferle:2004, Schori:2004}. We allowed for an additional hold time, keeping the sum of modulation and hold time constant at $200\,\text{ms}$.  To quantify the response, we adiabatically increased the lattice depth to reach the atomic limit ($j\approx 0$) and measured the temperature of the system with a recently developed scheme based on single-atom-resolved detection \mycite{Sherson:2010}. It is the high sensitivity of this method which allowed us to reduce the modulation amplitude by almost an order of magnitude compared to earlier experiments \mycite{Stoeferle:2004, Schori:2004} and to stay well within the linear response regime (see Supplementary Information). 

 \begin{figure*}
    \centering
   \includegraphics[width=1.5\columnwidth]{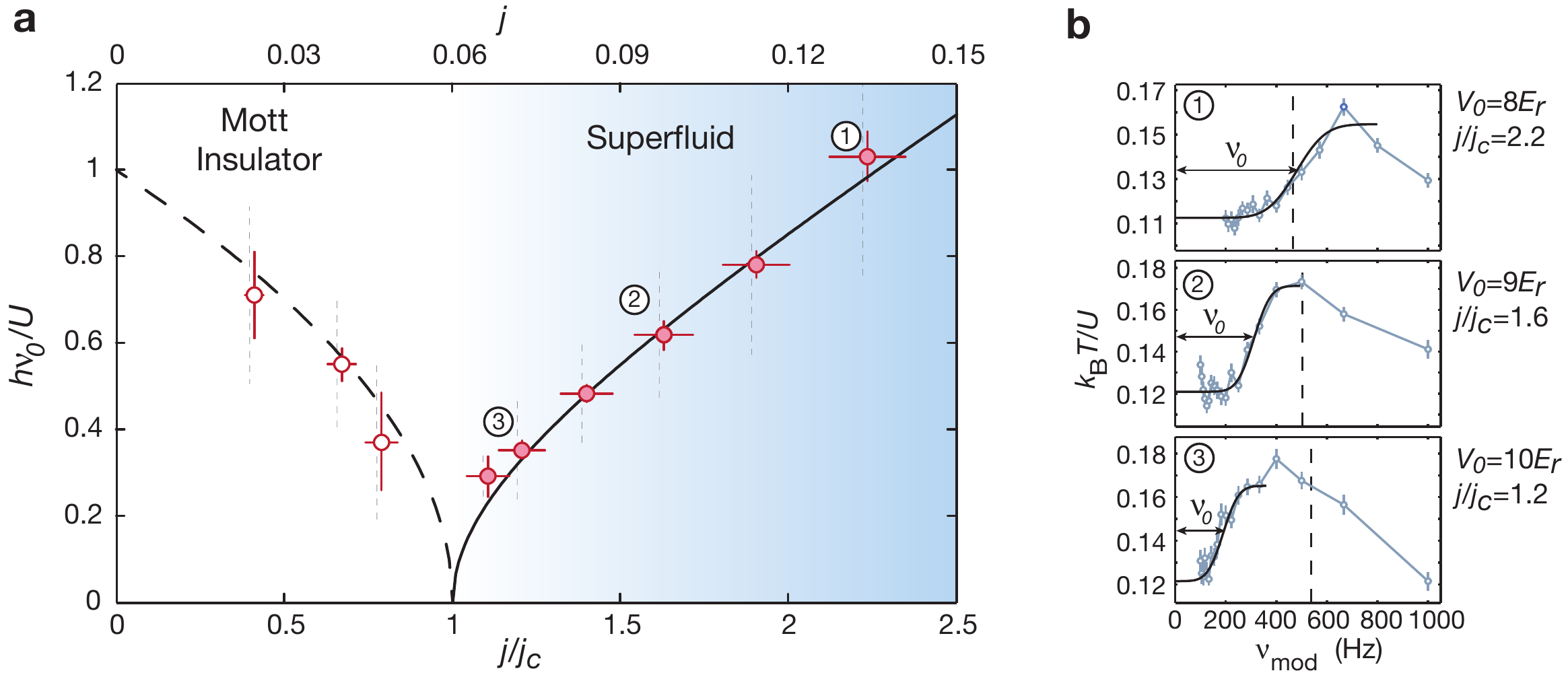}
    \caption{{\bf Softening of the Higgs mode.} \textbf{a}, The fitted gap values $h \nu_0/U$ (circles) show a characteristic softening close to the critical point in quantitative agreement with  analytic predictions for the Higgs and the Mott gap (solid line and dashed line, see text). Horizontal and vertical errorbars denote the experimental uncertainty of the lattice depths and the fit error for the center frequency of the error function, respectively (see Methods). Vertical dashed lines denote the width of the fitted error function \change{and characterize the sharpness of the spectral onset}. The blue shading highlights the superfluid region. \textbf{b}, Temperature response to lattice modulation (circles and connecting blue line) and fit with an error function (solid black line) for three different points in a, labeled by corresponding numbers. With the coupling $j$ approaching the critical value $j_c$, the change of the gap values to lower frequencies  is clearly visible (from panel 1 to 3). Vertical dashed lines mark the frequency $U/h$ corresponding to the on-site interaction. Each data point results from an average of the temperatures over $\approx 50$ experimental runs. Error bars represent the $1\sigma$ standard deviation.}
  \end{figure*}
The results for selected lattice depths $V_0$ are shown in Fig.\,2b. We observe a gapped response with an asymmetric overall shape that will be analysed in the following paragraphs. Notably, the maximum observed temperature after modulation is well below the `melting' temperature for a Mott insulator in the atomic limit \mycite{Gerbier:2007} $T_{\text{melt}}\approx 0.2\,U/k_B$, demonstrating that our experiments probe the quantum gas in \change{the} degenerate regime. To obtain numerical values for \change{the onset of spectral response}, we fitted each spectrum with an error function centred at a frequency $\nu_0$ (solid black lines in Fig.\,2b).  With $j$ approaching $j_c$, the shift of the gap to lower frequencies is already visible in the raw data (Fig.\,2b and Fig.\,5a) and becomes even more apparent for the fitted gap $\nu_0$ as a function of $j/j_c$ (Fig.\,2a, filled circles). The $\nu_0$ values are in quantitative agreement with a prediction for the Higgs gap at commensurate filling $h\nu_{\rm{SF}}/U=[(3\sqrt{2}-4)(1+j/j_c)]^{1/2}(j/j_c-1)^{1/2}$ (solid line) based on an analysis of variations around a mean field state \mycite{Huber:2007,Altman:2002} (throughout the manuscript, we rescaled $j_c$ in the theoretical calculations to match the value $j_c\simeq 0.06$ obtained from Quantum Monte-Carlo simulations \mycite{Capogrosso:2008}).

 \change{The sharpness of the spectral onset can be quantified by the width of the fitted error function, which is shown as vertical dashed lines in Fig.\,2a. Approaching the critical point, the spectral onset becomes sharper, while the width normalized to the center frequency $\nu_0$ remains constant (see Supplementary Information Fig.\,8). The latter indicates that the width of the spectral onset scales in the same way with the distance to the critical point as the gap frequency.}

We observe similar gapped responses in the Mott insulating regime (see Supplementary Information and Fig.\,5a), with the gap closing continuously when approaching the critical point (Fig. 2a, open circles). We interpret this as a result of \change{combined particle and hole excitations} with a frequency given by the Mott excitation gap that closes at the transition point \mycite{Huber:2007}. The fitted gaps are consistent with the Mott gap $h\nu_{\rm{MI}}/U=[1+(12\sqrt{2}-17)j/j_c]^{1/2}(1-j/j_c)^{1/2}$ predicted by mean field theory \mycite{Huber:2007} (dashed line).

\begin{figure}[t]
    \centering
   \includegraphics[width=1\columnwidth]{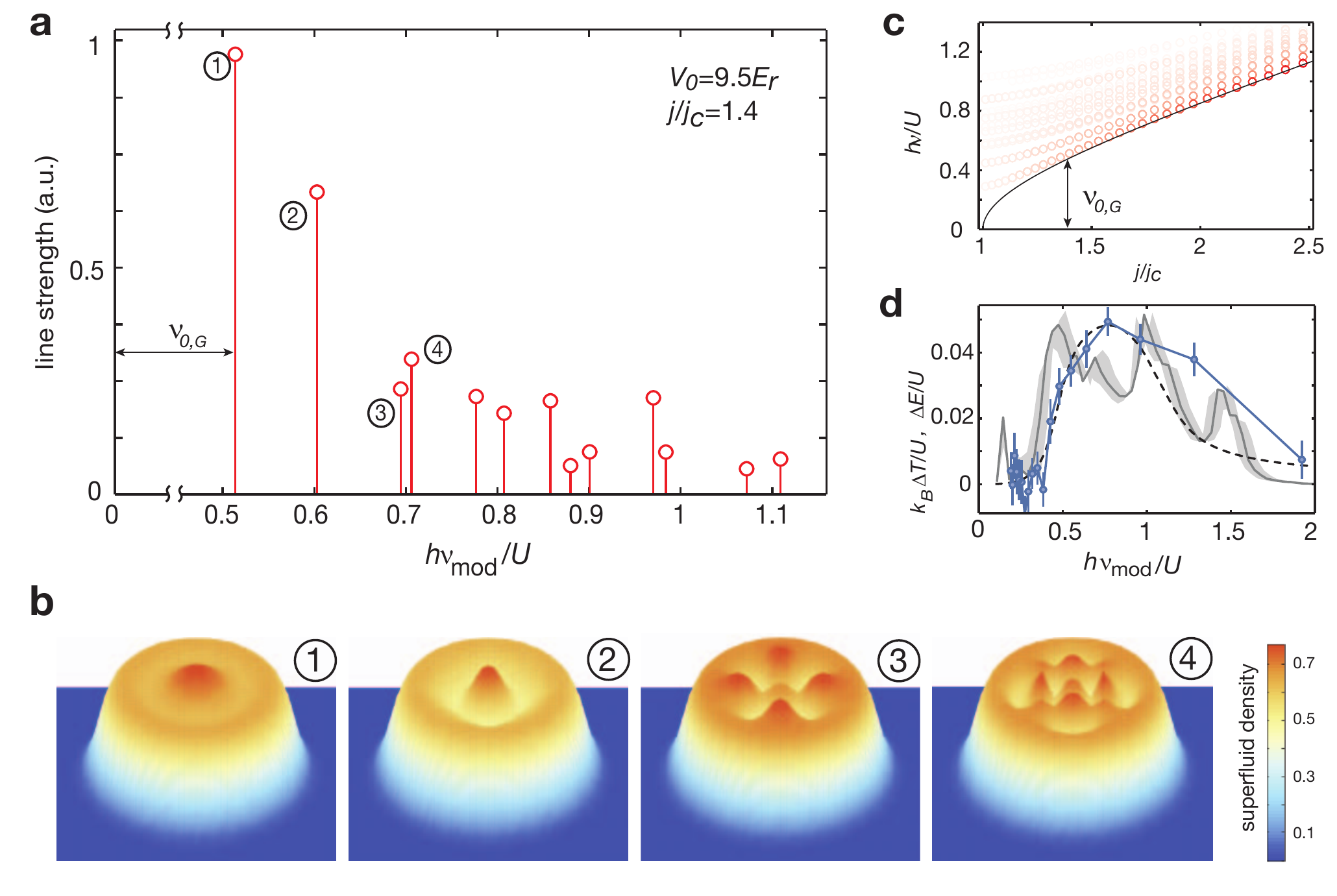}
    \caption{{\bf Theory for in-trap response.} \textbf{a}, A diagonalization of the trapped system in a Gutzwiller approximation shows a discrete spectrum of amplitude-like eigenmodes. Shown on the vertical axis is the strength of the response to a modulation of $j$.  Eigenmodes of phase-type are not shown (see Methods) \change{and $\nu_{0,G}$ denotes the gap as calculated in the Gutzwiller approximation.} \textbf{b}, In-trap superfluid density distribution for the four amplitude modes with lowest frequency marked by corresponding numbers in a. In contrast to the superfluid density, the total density of the system stays almost constant (not shown). \textbf{c}, Discrete amplitude mode spectrum for various couplings $j/j_c$. Each red circle corresponds to a single eigenmode with the intensity of the color being proportional the line strength. The gap frequency of the lowest-lying mode follows the prediction for commensurate filling (solid line, same as in Fig.\,2a) until a rounding off takes place close to the critical point due to the finite size of the system. \textbf{d}, Comparison of the experimental response at $V_0=9.5E_r$ (blue circles and connecting blue line) with a $2\times 2$ cluster mean field simulation (gray line and shaded area) and a heuristic model (dashed line, for details see text and Methods). The simulation was done for $V_0=9.5E_r$ (gray line) and for $V_0=(1\pm2\%)\cdot 9.5E_r$ (shaded gray area) in order to account for the experimental uncertainty of the lattice depth and predicts the energy absorption per particle $\Delta E$. }
  \end{figure}
The observed softening of the \change{onset of spectral response} in the superfluid regime has lead to \change{an} identification with collective excitations of Higgs type. To gain further insight into the full in-trap response, we \change{calculated} the eigenspectrum of the system in a Gutzwiller approach \mycite{Huber:2007, Bissbort:2011} (see Methods  and Supplementary Information). The result is a series of discrete eigenfrequencies (Fig.\,3a) and the corresponding eigenmodes show in-trap superfluid density distributions, which are reminiscent of the vibrational modes of a drum (Fig.\,3b). The frequency of the lowest-lying amplitude-like eigenmode closely follows the long-wavelength prediction for homogeneous commensurate filling $\nu_{SF}$ over a wide range of couplings $j/j_c$ until the response rounds off in the vicinity of the critical point due to the finite size of the system (Fig.\,3c). Fitting the low-frequency edge of the experimental data can be interpreted as extracting the frequency of this mode, which explains the good quantitative agreement with the prediction for the homogeneous commensurate filling in Fig.\,2a. \change{Modes at different frequencies from the lowest-lying amplitude-like mode broaden the spectrum only above the onset of spectral response.} 

An eigenmode analysis, however, does not yield any information about the finite spectral width of the modes\change{, which stems from the interaction between amplitude and phase excitations}. We will consider the question of the spectral width by analysing the low-, intermediate- and high-frequency part of the response separately. We begin by examining the low-frequency part of the response, \change{which is expected to be governed by a process coupling a virtually excited amplitude mode to a pair of phase modes with opposite momentum. As a result, the response of a strongly interacting two-dimensional superfluid is expected to diverge at low frequencies, if the probe in use couples longitudinally to the order parameter \mycite{Sachdev:1999,Zwerger:2004,Sachdev:2011} (e.g., to the real part of $\Psi$, if $\Psi$ was chosen along the real axis), as it is the case for neutron scattering. If, instead, the coupling occurs in a rotationally invariant fashion (i.e.,to $|\Psi|^2$), as expected for lattice modulation, such a divergence could be avoided and the response is expected to scale with $\omega^3$ at low frequencies \mycite{Chubukov:1994,Huber:2008,Podolsky:2011}.} Combining this result with the scaling dimensions of the response function \change{for a rotationally symmetric perturbation}, we expect the low-frequency response to be proportional to $(1-j/j_c)^{-2} \omega^3$ (see Methods). The experimentally observed \change{signal} is consistent with \change{this} scaling at the `base' of the absorption feature (Fig.\,4). This indicates that the low-frequency part is dominated by only a few in-trap eigenmodes, which approximately show the generic scaling of the homogeneous system \change{for a response function describing coupling to $|\Psi|^2$}.

 \begin{figure}[t]
    \centering
   \includegraphics[width=0.8\columnwidth]{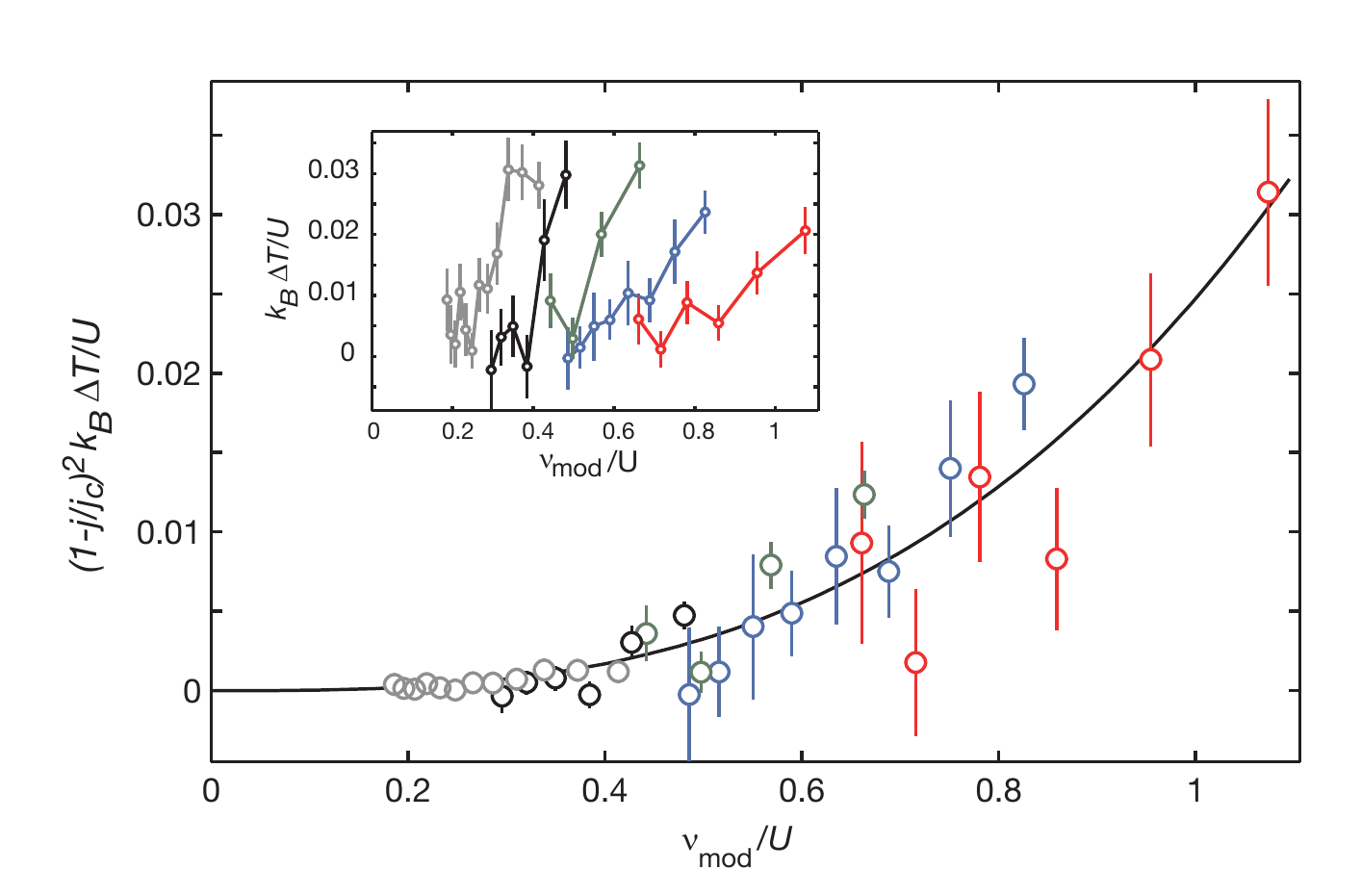}
    \caption{{\bf Scaling of the low-frequency response.} The low-frequency response in the superfluid regime shows a scaling compatible with the prediction $(1-j/j_c)^{-2}\omega^3$ (see Methods). Shown is the temperature response rescaled with $(1-j/j_c)^2$ for $V_0=10 E_r$ (grey), $9.5E_r$ (black), $9E_r$ (green), $8.5E_r$ (blue),  $8E_r$ (red) as a function of the modulation frequency. The black line is a fit of the form $a\, \omega^b$, with a fitted exponent $b=2.9(5)$. The inset shows the same data points without rescaling for comparison. }
  \end{figure}
In the intermediate-frequency regime, it remains a challenge to construct a first-principle analytical treatment of the in-trap system including all relevant decay and coupling processes. Lacking such a theory, we constructed a heuristic model combining the discrete spectrum from the Gutzwiller approach (Fig.\,3a) with the lineshape for a homogeneous system based on an $O(N)$ field theory in two dimensions, \change{calculated in the large $N$ limit} \mycite{Chubukov:1994,Podolsky:2011} (see Methods). An implicit assumption of this approach is a continuum of phase modes, which is approximately valid in our case because the frequency spacing between different phase modes is much smaller than the typical gap to the lowest amplitude mode. The model yields quantitative agreement with the low- to intermediate-frequency experimental data for a range of couplings (dashed black line in Fig.\,3d and Supplementary Information), where a relativistic field theoretical treatment of this type is applicable. Further, the response at frequencies higher than twice the absorption edge remains slightly underestimated.

\begin{figure}
    \centering
    \includegraphics[width=0.9\columnwidth]{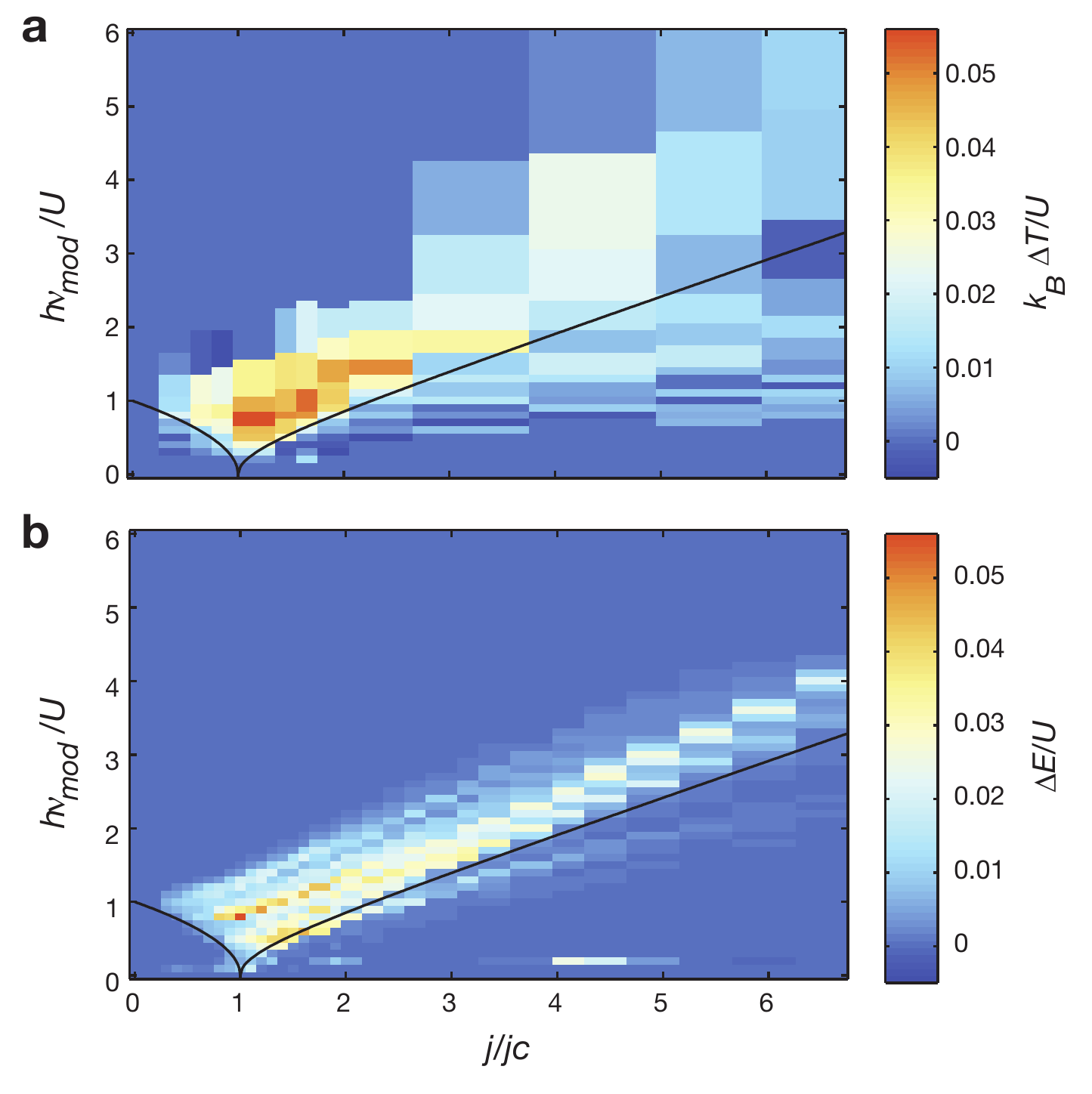}
    \caption{{\bf Response from the strongly to the weakly interacting limit.} \textbf{a}, Change in Temperature $\Delta T$ as a function of $j/j_c$ and the modulation frequency $\nu_{\rm{mod}}$. A pronounced feature close to $j/j_c=1$ directly shows the existence of the gap and its softening. Approaching the weakly interacting limit (higher $j/j_c$), the response broadens and vanishes. \textbf{b}, Simulation using a variational  $2\times 2$ cluster wave function predicting the energy absorption per particle $\Delta E$ for the same parameter range. The simulation shows agreement with the experimental data near the critical point in both the softening of the response and the overall width of the absorption band. However, the simulation does not fully reproduce the vanishing of the response at higher $j/j_c$ values. A splitting in the excitation structure at $j/j_c\approx3$ is visible, which might also be present in the experimental data. A low-frequency feature associated with density oscillations at the edges of the trap \change{due to the excitation of phase-like modes} is clearly seen in the simulations. This feature occurs below the lowest measured frequency in the experiment and thus is not visible in a, (except in the vicinity of the critical point, where the lowest modulation frequencies are close to this feature). Black solid lines show the mean field predictions as plotted in Fig.\,2a.}
\end{figure}
Part of this high-frequency response might stem from the excitation of several amplitude modes or combinations of amplitude and phase modes, which cannot be described with the Gutzwiller approximation used so far and \change{is only partly captured in the field theoretical treatment}. Therefore, we performed a dynamical simulation based on a  $2\times2$ cluster variational wave function, which captures the excitation of multiple modes as well as inter-mode coupling, at least at high momenta. The result is compared with experimental data in Fig.\,3d and shows good overall agreement (also compare Fig.\,5a and 5b near the critical point). Notably, the simulation predicts the low-frequency edge, the overall width and the absolute strength of the experimental signal without any fitting parameters.  The simulation, which also accounts for fluctuations of the experimental parameters, shows a fine structure which is not observed in the experiment. This indicates that the $2\times2$ cluster treatment still cannot fully capture the broadening of the modes due to coupling with low-energy phase modes.

Our analysis, so far, has shown the existence of an amplitude mode in the Bose-Hubbard model close to the critical point ($j/j_c\approx 1$), where the low-energy description of the system is approximately Lorentz invariant. In the weakly interacting limit ($j/j_c\gg1$), however, the low-energy description (Gross-Pitaevski theory) becomes effectively Galilean invariant, which forbids \change{the existence of} such a mode \mycite{HuberPhd:2008,Varma:2002}. To probe the evolution of the amplitude-mode response when approaching the weakly interacting limit, we extended our measurements to higher values of the coupling $j$. The results are shown in Fig.\,5a as a density plot, where a pronounced signal for $j/j_c\lesssim3$ directly shows the softening of the mode close to the critical point. Approaching the weakly interacting limit with higher $j/j_c$ values, the response gradually broadens and finally disappears. Despite earlier theoretical treatments of the system in this regime  \mycite{Huber:2008,Cazalilla:2006,Huber:2011}, a prediction of the  disappearance of the response is still lacking. Also, results from the $2\times2$ cluster variational wave function approximation could only partially capture this effect (Fig.\,5b).

In conclusion, we could identify and study long-wavelength Higgs modes in a neutral two-dimensional superfluid close to the quantum phase transition to a Mott insulating state. This was enabled by recent advances in the high-resolution imaging of single atoms in optical lattices \mycite{Bakr:2010, Sherson:2010}, leading to a new level of precision for the spectroscopy of ultracold quantum gases. The obtained spectra show softening at the quantum phase transition and \change{are consistent with the generic $\omega^3$ low-frequency scaling for a rotationally invariant coupling to the order parameter in a two-dimensional strongly interacting superfluid}. Furthermore, our results challenge the development of a quantitative theory valid between the strongly and the weakly interacting regimes capable of predicting the observed disappearance of the response. \change{Our data also} call for a first-principle treatment of the discrete nature of Higgs modes in a confined system. In this regard, we note an interesting connection to particle physics, where the Higgs boson spectrum within the conjectured compact extra dimensions \mycite{Randall:1999} may acquire a similar discrete spectrum.

\section*{Acknowledgements}
  We thank C. Weitenberg and J. F. Sherson for their
contribution to the design and construction of the apparatus. We thank Daniel Podolsky, Wilhelm Zwerger, Subir Sachdev, Rajdeep Sensarma, Walter Hofstetter, Ulf Bissbort, Lode Pollet and Nikolay Prokof'ev for helpful discussions. We acknowledge funding by MPG, DFG, EU (NAMEQUAM, AQUTE, Marie Curie
Fellowship to M.C.), JSPS (Postdoctoral Fellowship for Research Abroad to T.F.) and California Institute of Technology (Lee A. DuBridge fellowship for D.P.).

\bibliography{higgs_mode_bibliography}

\section*{Methods}
 \textbf{Experimental details.} The preparation of the two-dimensional degenerate gas is described in Ref. \mycite{Endres:2011}. During the experiment, the gas was held in a single anti-node of a vertical optical standing wave with a depth of $20(2)E_r$, where $E_r$ denotes the lattice recoil energy $E_{r}=h^2/(8m a_{\rm lat}^2)$ with $m$ the atomic mass of $^{87}$Rb.  The lattice constant for the vertical and both horizontal optical lattices was $a_{\rm lat}=532\,\rm{nm}$ and the trapping frequencies for the two-dimensional system were typically $60\,\rm{Hz}$. The ramp for lattice loading and the ramp to the atomic limit were s-shaped with a total duration of $120\,\rm{ms}$ and $75\,\rm{ms}$ respectively.   Our systems contained an atom number of $190(36)$ resulting in a central density close to one atom per lattice site. The data point in Fig.\,2a at $j/j_c\approx 1.1$ was taken with slightly different parameters ($T_{\rm{tot}}=300\,\rm{ms}$ and $T_{\rm{mod}}=15\, \tau$, see Fig.\,1b) and is excluded in Fig.\,4 and Fig.\,5a.

\textbf{Calibration of the lattice depths.} We calibrated the lattice depths by performing amplitude modulation spectroscopy \mycite{Endres:2011} with an estimated calibration uncertainty of 1\%. To minimize drifts of the lattice depths, we typically repeated this calibration for the horizontal lattice axes after $80$ experimental runs. We observed a drift of the lattice depths between these calibrations of maximally $2\%$, but in most cases no change was observable.

\textbf{Determination of the Bose-Hubbard parameters.}
The Bose-Hubbard Hamiltonian is given by
\begin{align}
H_{\text{\rm{BH}}}=-J \sum_{\langle i k \rangle} \hat{b}^\dagger_i \hat{b}_k+\frac{U}{2}\sum_{i} \hat{n}_i (\hat{n}_i-1)+\sum_i(V_i-\mu) \hat{n}_i,
\end{align}
where $\hat{b}^\dagger_i$ ($\hat{b}_i$) is the boson creation (annihilation) operator on lattice site $i$, $\hat{n}_i$ is the boson number operator, $J$ is the hopping matrix element, $U$ is the on-site interaction, $\mu$ is the chemical potential, and $V_i$ describes the harmonic trapping potential.  The Bose-Hubbard parameters $J$ and $U$ were calculated from the lattice depths by a numerical band-structure calculation \mycite{Bloch:2008}. The uncertainties of the coupling $j=J/U$, stated as horizontal errorbars in Fig.\,1a, result from the experimental uncertainty of the lattice depths of about $2\%$.

\textbf{Fit of the mode gap.} 
We fitted the temperature response with $ T=T_0+\Delta T/2\,\{\mbox{erf}[\frac{1}{\sigma_e}(\nu_{\rm{mod}}-\nu_0)]+1\}$, \change{ where $\mbox{erf}(x)$ denotes the error function.} The fitting parameters were the temperature offset $T_0$, the temperature increase $\Delta T$, the width $\sigma_e$ and the center frequency $\nu_0$. \change{The fit function is a model free approach to extract numerical values for the onset of spectral response. The center frequency $\nu_0$ (circles in Fig. 2a) is a measure for the position of the spectral onset, while the width $\sigma_e$ (vertical dashed lines in Fig. 2a) is a measure for its sharpness. The width $\sigma_e$ can also be seen as an estimation for the maximum error on the extraction of the position of the onset.} During the least-square optimization, data points at frequencies larger than $\nu_0+2.5\,\sigma_e$ were excluded. The vertical errorbar in Fig.\,2a is given by the $1\sigma\,$fitting error for $\nu_o$ and the vertical dashed lines denote $\pm\sigma_e$.

\textbf{Scaling of the low-frequency response.}
The amplitude mode response at low frequencies is expected to be proportional to $\omega^3$, which is observed in a weak-coupling expansion \mycite{Podolsky:2011}, in a large $N$ expansion \mycite{Chubukov:1994,Podolsky:2011} and in the quantum phase model \mycite{Huber:2008}. Additionally, dimensional analysis shows that the amplitude mode response for an $O(N)$ field theory should follow a scaling of the form \mycite{Sachdev:2011,Podolsky:2011} $F(\omega,\frac{j}{j_c})=A\Delta^{3-2/\nu}\Phi(\frac{\omega}{\Delta})$, where $\Delta\propto(1-\frac{j}{j_c})^\nu$ is a typical energy scale, $\nu$ is the critical exponent associated with $\Delta$, $A$ is a constant and $\Phi$ a universal function. Combining this scaling with the $\omega^3$ prediction yields $F(\omega,\frac{j}{j_c})=A(1-\frac{j}{j_c})^{-2}\omega^3$ at low frequencies. For the plot in Fig.\,4, we chose frequencies in a span from $\nu_0-1.5\,\sigma_e$ to $\nu_0+0.5\,\sigma_e$, with $\nu_0$ and $\sigma_e$ taken from the error function fit of the individual responses. 

{\bf Gutzwiller calculation of the Eigenmodes in a trap.}
To perform the eigenmode analysis, we used the Gutzwiller trial wave function \mycite{Huber:2007, Bissbort:2011}
\begin{align}
|\Psi_{1\times1}\rangle &=e^{i \phi} \prod_i ( \alpha_i(t) \left |0 \right\rangle_i\nonumber \\& + \sqrt{1-|\alpha_i(t)|^2 -|\gamma_i(t)|^2} |1\rangle_i  + \gamma _i (t) |2\rangle_i),
\end{align}
where $\alpha_i(t)$ and $\gamma_i(t)$ are variational parameters, $|n\rangle_i$ corresponds to a state with $n$ bosons on site $i$, and $\phi$ is an overall phase. First, we obtained the stationary solution $|\Psi_{1\times1}^0\rangle$ (corresponding to $\{\alpha^0_i, \, \gamma^0_i\}$) by minimizing $\langle \Psi^0_{1\times1} | H_{\text{BH}} | \Psi^0_{1\times1} \rangle$ in the entire trap. Next, we linearized the equations of motion, which were obtained by minimizing the effective action $\langle \Psi_{1\times1} |i\partial_t -H_{\rm{BH}} | \Psi_{1\times1}\rangle$ around the stationary solution. The resulting eigenvalue problem was solved by a Bogoliubov transformation $M_{k,ir}$ that relates Bogoliubov creation $f_k^\dagger$ (and annihilation $f_k$) operators to the small fluctuations $\delta\alpha_i$ and $\delta\gamma_i$ around the stationary solution
\begin{align}
f_k^\dagger=\sum_{i} \left(M_{k,i1} \delta\alpha_i + M_{k,i2} \delta\alpha_i^*
+ M_{k,i3} \delta\gamma_i + M_{k,i4} \delta\gamma_i^*\right),
\end{align}
where $k$ is the eigenmode index.  \change{We can identify the modes as amplitude-like or phase-like using a measure of  `amplitudeness'}
\begin{align}
A=\sum_i M_{k,i1} M_{k,i3},
\end{align}
which is positive for amplitude-like modes and negative for phase-like modes (see Supplementary Information). 

To describe lattice modulation spectroscopy, we separated the Bose-Hubbard Hamiltonian into a time independent part that describes the system with no modulation and a time dependent part that describes the lattice modulation: $H_{\rm{BH}}=H_0+\sin(\omega t) H'$. The rate of excitation of the $k$-th amplitude or phase mode is given by Fermi's golden rule:
$\Gamma(\omega)=\delta(\omega-\omega_k) |\langle  \Psi^0_{1\times1}  f_k | H' |  \Psi^0_{1\times1} \rangle|^2$. The line strengths plotted in Fig. 3a are proportional to $ S_i=|\langle  \Psi^0_{1\times1}  f_k | H' |  \Psi^0_{1\times1} \rangle|^2$. \change{The line strengths decrease with increasing frequency, because higher energy modes show short wavelength spatial variations and do not efficiently couple to lattice modulation.}

\textbf{Heuristic model.} We constructed a heuristic model (dashed line in Fig.\,3d and Supplementary Information), which combines the frequencies and line strengths of our Gutzwiller calculation with the shape of the response calculated by field theoretical methods. For a given $j/j_c$ value, the Gutzwiller approach yields a series of amplitude-like normal modes with frequencies $\nu_i$ and corresponding line strengths $S_i$. The heuristic model consists of summing up a response function $F(\nu_i,\nu_{\rm{mod}})$ for each of this frequencies weighted with the corresponding line strengths. A calculation based on a large $N$ expansion of a two-dimensional $O(N)$ field theory \mycite{Chubukov:1994,Podolsky:2011} yielded a scalar response function for the homogeneous and commensurate system of the form
\begin{align}
F(\nu,\nu_{\rm{mod}})\propto \frac{\nu_{\rm{mod}}^3}{(\nu_{\rm{mod}}^2-\nu^2)^2+4\gamma^2\nu_{\rm{mod}}^2}.
\end{align}
A parametrization of the $N=2$ case of the model can be found in Refs. \mycite{Altman:2002,Polkovnikov:2005} and yields $h \gamma/U=\frac{1}{8}$. Assuming this response function at each individual normal mode (and measuring all frequencies in units of $U/h$) results in the final model function\\
\begin{align}
F_h(\nu_{\rm{mod}})=A_1+A_2\sum_i S_i \frac{\nu_{\rm{mod}}^3}{(\nu_{\rm{mod}}^2-\nu_i^2)^2+4\gamma^2\nu_{\rm{mod}}^2},
\end{align}
with $\gamma=\frac{1}{8}$ and fit parameters $A_1$ and $A_2$. \\
{\bf Dynamical evolution: $2\times2$ cluster wave functions.} We performed a study of the dynamical evolution of the system using $2\times2$ cluster variational wave functions
\begin{widetext}
\begin{align}
|\Psi_{2\times2}\rangle=\prod_{i} \left[
 a_i(t) \left|
\begin{array}{cc}
0 & 0\\
0 & 0
\end{array}
\right\rangle 
+
b_i(t) \left|
\begin{array}{cc}
1 & 0\\
0 & 0
\end{array}
\right\rangle 
 +c_i(t) \left|
\begin{array}{cc}
0 & 1\\
0 & 0
\end{array}
\right\rangle 
+\dots
\right],
\end{align}
\end{widetext}
where $a_i(t)$, $b_i(t)$, $c_i(t)$, \ldots are the variational parameters. \change{We restrict the maximum occupation number per site to two}. To initialize the dynamics, we obtained the initial trial wave function, corresponding to the state of the system before modulation spectroscopy begins, by minimizing $\langle \Psi_{2\times2}|H_{\rm{BH}}|\Psi_{2\times2}\rangle$. Next, we dynamically evolved the trial wave function during the modulation drive, the hold time and the ramp to the atomic limit (see Fig 1b). Finally, we measured the total energy absorption per particle $\Delta E$ of the resulting state (in units of the on-site interaction $U$ in the atomic limit).

\section*{Supplementary Information}
\subsection{Raw data, linear response and width of the model function}
In Fig.\,6 we show the raw data for all data points plotted in Fig.\,2a of the main text. The closing of the gap coming from the superfluid side of the transition ($j/j_c>1$) is clearly visible and is followed by a reopening in the Mott insulating regime ($j/j_c<1$).\\
We probed the time dependence of the response for two different combinations of lattice depth $V_0$ and modulation frequency $\nu_{\rm{mod}}$. The results are shown in Fig. 7. A linear response as a function of the number of oscillation cycles is visible up to $40$ modulation cycles. Notably, our experiments were performed at $20$ cycles, staying well within the regime of linear temperature response.  A linear fit yielded a slope of $1.8(2) \cdot 10^{-3} \frac{k_B T}{U}\frac{1}{\tau}$ in both cases, where $\tau$ is the time for a single cycle of the modulation.

\begin{figure}[h!]
    \centering
   \includegraphics[width=0.95\columnwidth]{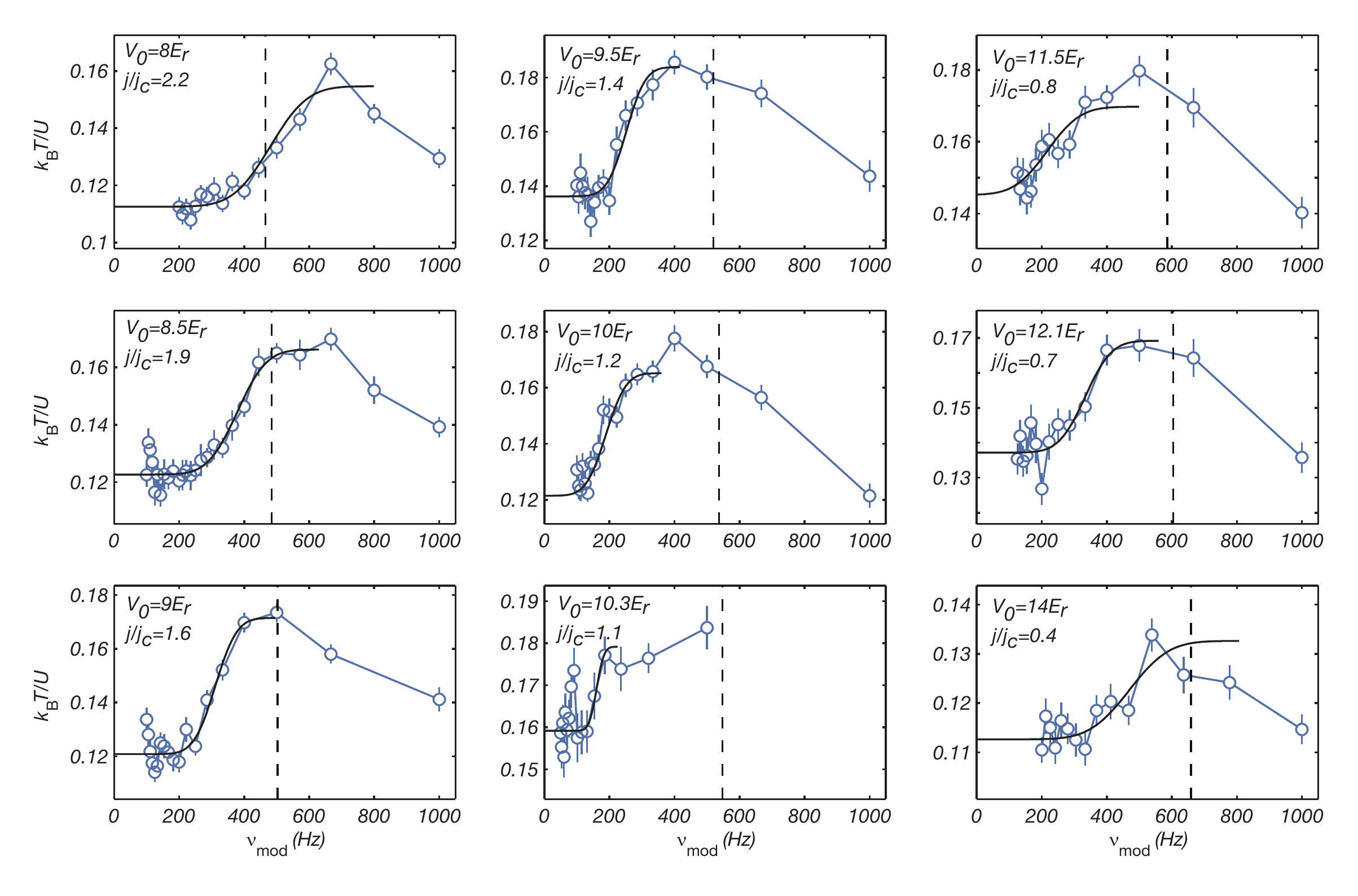}
    \caption{{\bf Raw data and fits for all data points shown in Figure 2a of the main text.} Each data point results from an average of the temperatures from about $50$ experimental runs. Error bars represent the standard deviation.}
\end{figure}  
\begin{figure}[h!t]
    \centering
   \includegraphics[width=0.6\columnwidth]{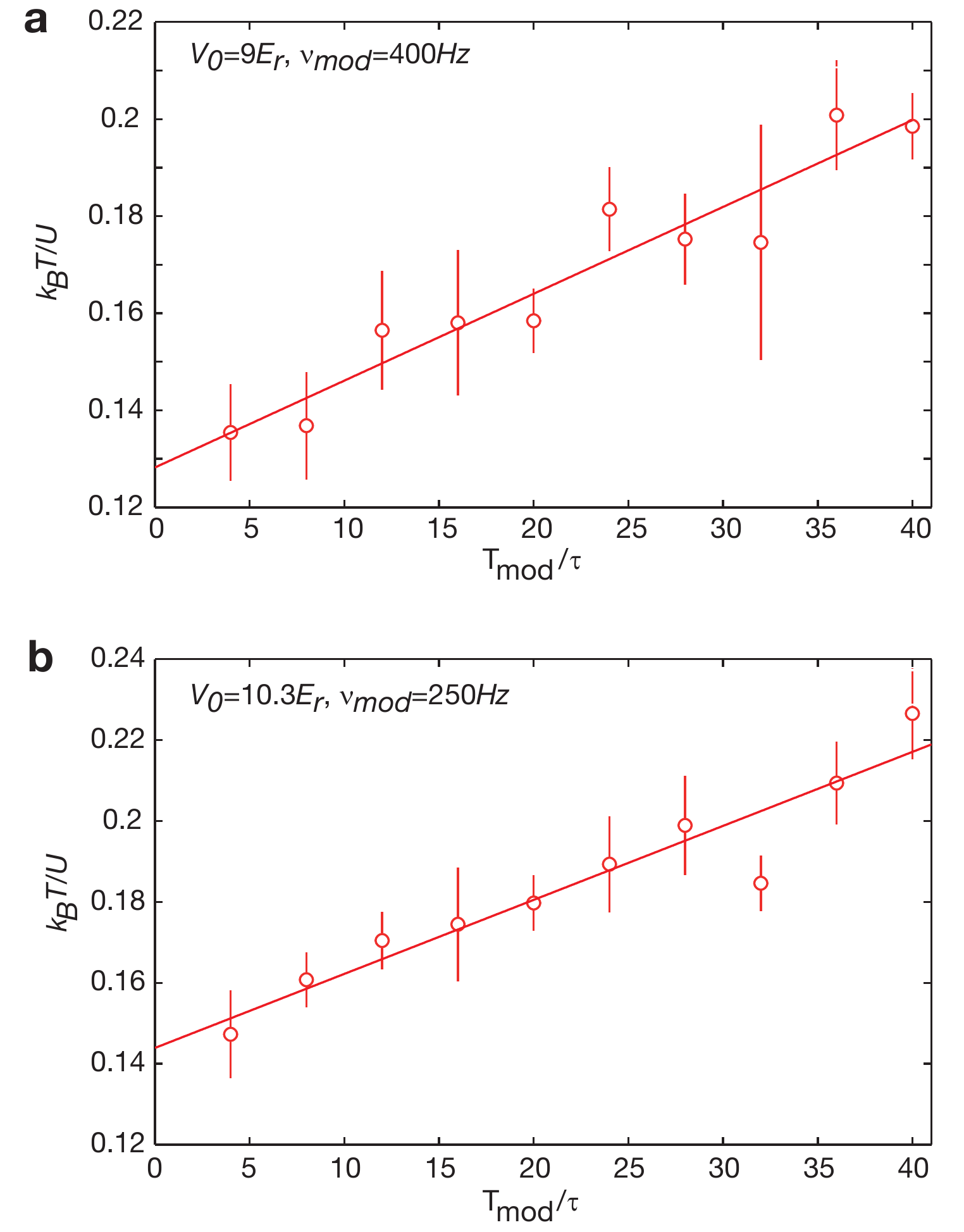}
    \caption{\textbf{Linear Response.}{ Temperature as a function of the number of modulation cycles for \textbf{a}, $V_0=9 E_r$ and $\nu_{\rm{mod}}=400\,$Hz and \textbf{b}, $V_0=10.3 E_r$ and  $\nu_{\rm{mod}}=250\,$Hz. }}
\end{figure}  
\begin{figure}[h!t]
    \centering
   \includegraphics[width=0.6\columnwidth]{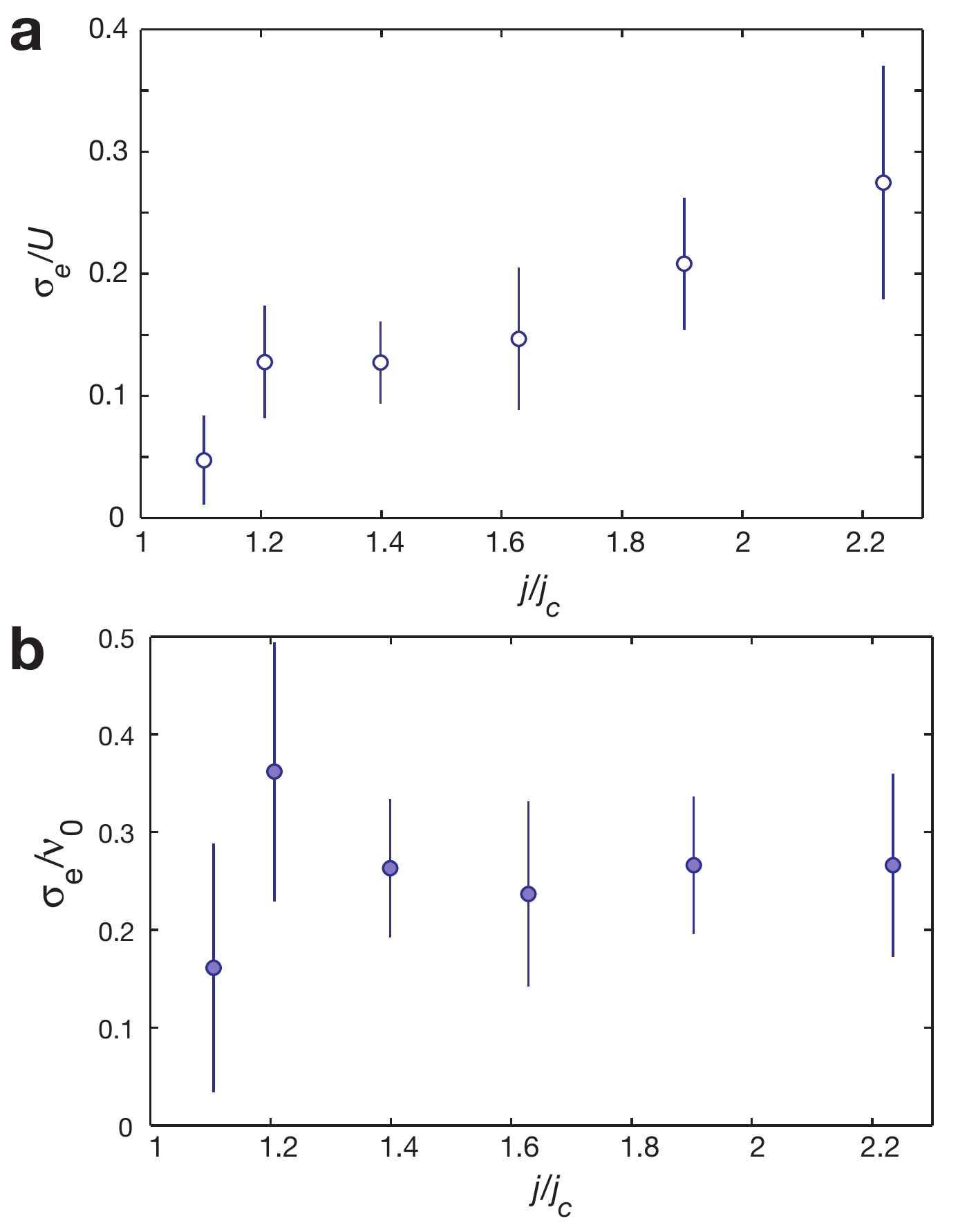}
    \caption{\textbf{Width of the model function.}{ \textbf{a}, Width $\sigma_e$ of the fitted error function for $j>j_c$. \textbf{b}, Width $\sigma_e$ divided by the center frequency $\nu_0$ of the fitted error function for $j>j_c$.}}
\end{figure}

\newpage
\change{
As shown in Fig.\,8a, the width $\sigma_e$ of the fitted error function decreases when approaching the critical point from the superfluid side ($j>j_c$) indicating that the sharpness of the spectral onset increases. However, the width $\sigma_e$ normalized to the center frequency of the error function $\nu_0$ remains constant within errorbars (see Fig.\,8b).}
\newpage
\subsection{Heuristic model in comparison with the experimental data}
In Fig.\,9 we show a comparison of the heuristic model (see Methods) with the experimental data in the superfluid region. We find good agreement for the low-frequency to mid-frequency response close to the phase transition, where the relativistic field-theoretical treatment is expected to be applicable.

\begin{figure}[h!]
    \centering
   \includegraphics[width=0.9\columnwidth]{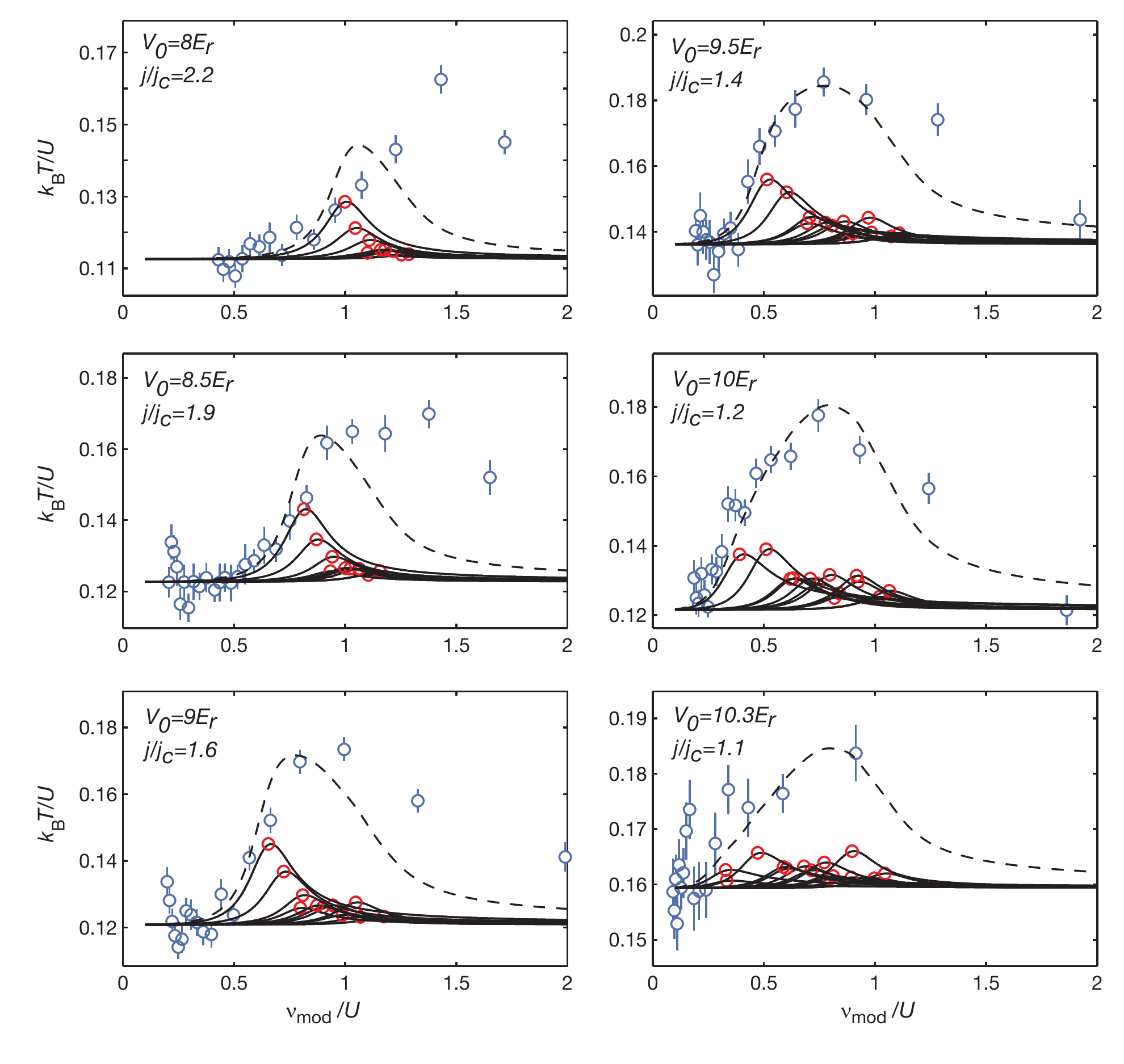}
    \caption{\textbf{Heuristic model.} Fit of the data in the superfluid regime (blue circles) with a heuristic model (dashed line). For details concerning the model see the Methods section. The individual contributions $A_2 S_i F(\nu_i,\nu_{\rm{mod}})$ are shown as solid lines. Red circles mark the frequency position $\nu_i$ of the corresponding normal mode.}
\end{figure} 

\subsection{Phase- and amplitude-like modes}
We verify that eigenmodes show amplitude-like or phase-like character by applying the amplitude measure, Eq.\,(4) of the Methods section. For each eigenmode, we compute the amplitude measure and plot it as a function of the eigenfrequency (see Fig.\,10a). The result shows two branches: a branch at lower frequencies with negative measure that corresponds to phase-like modes, and a gapped branch at higher frequencies with positive measure that corresponds to amplitude-like modes.\\
\change{Importantly, lattice modulation couples only to a few modes of the full spectrum (see Fig.\,10b). Most modes have strictly vanishing coupling strengths, because their spatial symmetry prevents an excitation with lattice modulation (where both axis are driven in phase and with the same amplitude). From the remaining modes, only the ones with the lowest frequencies of the respective branch show a significant coupling. At higher frequencies, the modes are dominated by short wavelength spatial variations avoiding an efficient coupling to (uniform) lattice modulation.

\begin{figure}
    \centering
   \includegraphics[width=0.7\columnwidth]{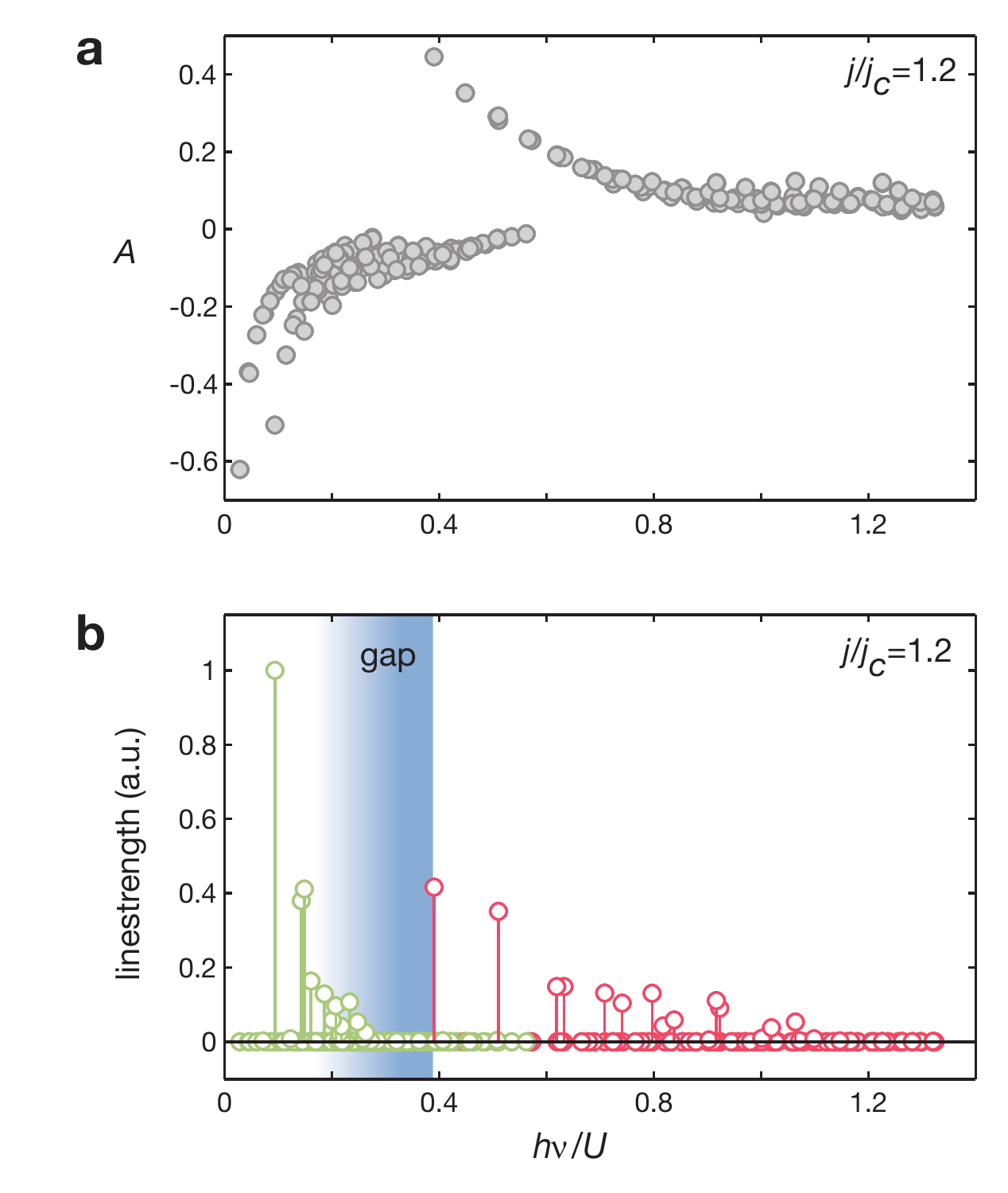}
    \caption{\textbf{Amplitude measure and line strength.} \change{\textbf{a,} Amplitude measure as a function of eigenfrequency for a trapped system at $j/j_c=1.2$. Amplitude-like (phase-like) modes have positive (negative) amplitude measure. \textbf{b,} Line strength for the same mode spectrum. Amplitude-like (phase-like) modes are shown in red (green). A large number of modes have strictly zero coupling strength due to their spatial symmetry. From the modes, which are allowed by symmetry, only the ones with long wavelength spatial variations (low-energy) couple significantly to lattice modulation. This leads to an effective gap between the amplitude-like and phase-like response (blue shading). The effective gap increases with the distance to the critical point (Fig.\,11).}}
\end{figure} 
\begin{figure}[t!]
    \centering
   \includegraphics[width=0.8\columnwidth]{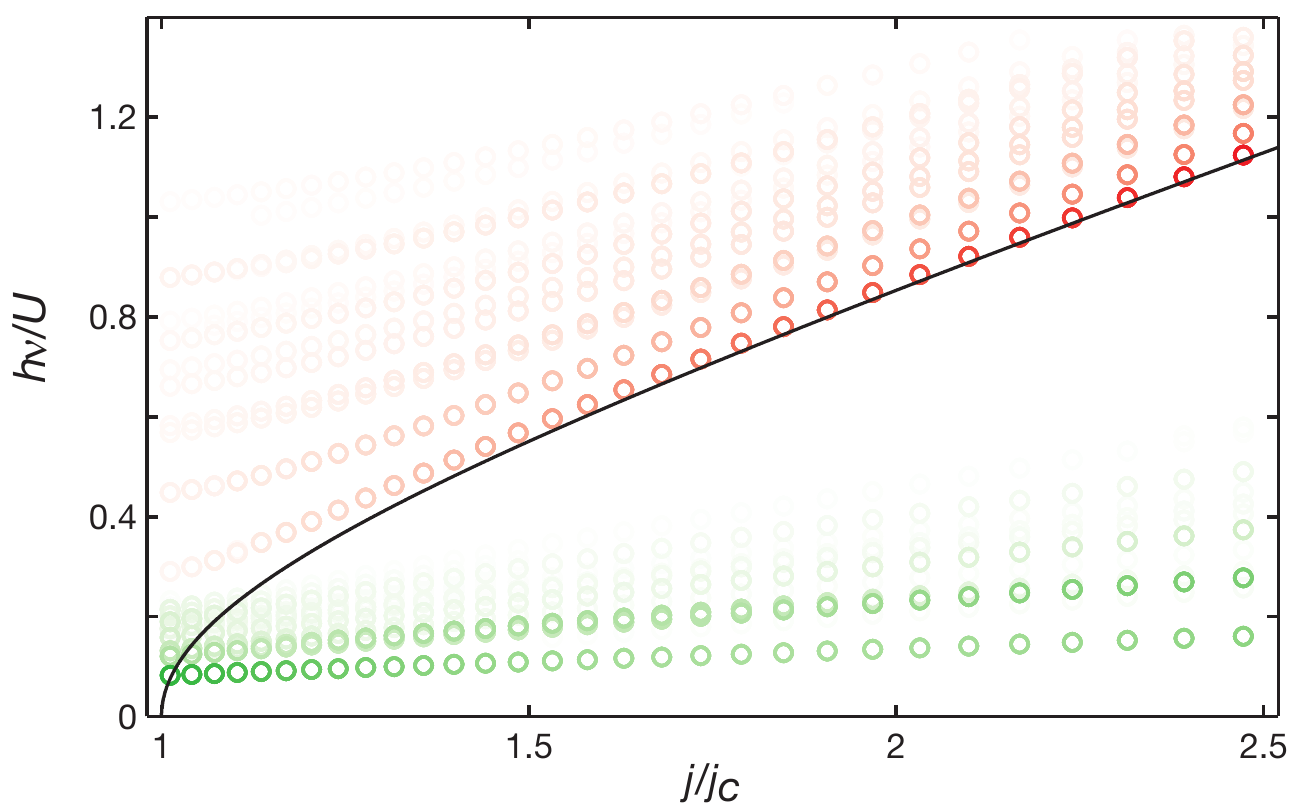}
    \caption{\textbf{Normal mode spectrum as a function of $j/j_c$.} \change{Each circle corresponds to an eigenmode with the intensity of the color being proportional to the line strength (see Methods). Amplitude-like (phase-like) modes are shown in red (green). The response from phase-like modes is only visible at low frequencies and separated from the amplitude-like response by a gap.  The experimental response in the main amplitude-mode peak is therefore expected to be unaltered by the presence of phase-like modes.}}
\end{figure} 
This results in an effective gap between the response from phase and amplitude-like modes (see Fig.\,10b and Fig.\,11). In the experiment, both responses are therefore expected to be well separated. In the spectra shown in Fig.\,9, a response from phase-like modes might be visible for the lowest experimental frequencies (e.g., for $V_0=9E_r$ at $\nu_{\rm mod}/U\approx0.2$ or for $V_0=10.3E_r$ at $\nu_{\rm mod}/U\approx0.15$). In all cases, these features are separated from the main peak and do not alter the analysis of the main response, except for the low-frequency scaling in Fig.\,4 of the main text, where the lowest frequency data points were excluded (see Methods). A response from phase-like modes is also visible for the cluster wave function data shown in Fig.\,5b of the main text as a sharp feature at very low frequencies (see also figure caption of Fig.\,5b), which is well separated from the main response.

\begin{figure}
    \centering
   \includegraphics[width=0.8\columnwidth]{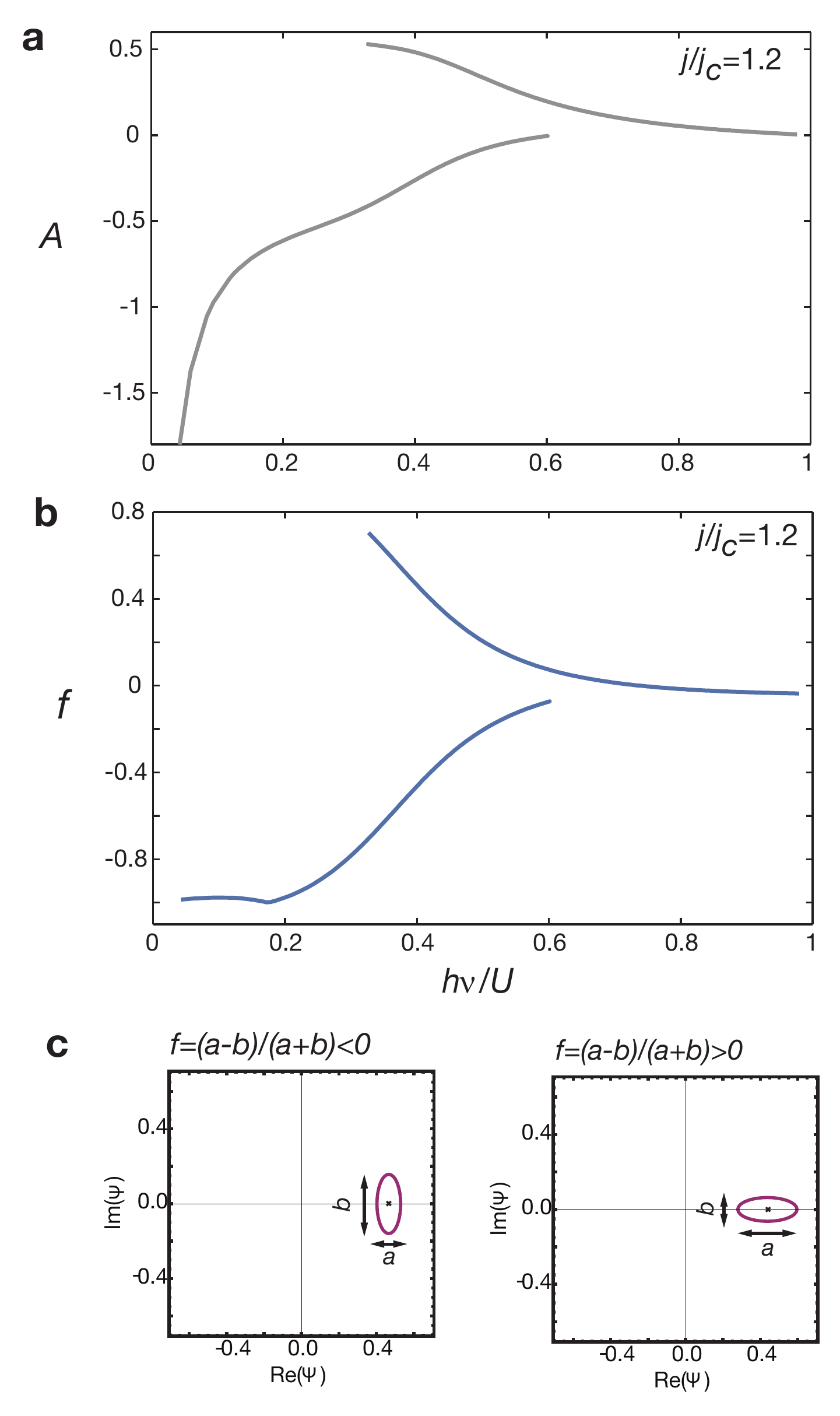}
    \caption{\textbf{Phase- and amplitude-like modes  for a homogeneous system.} \change{\textbf{a,} Amplitudeness measure $A$ for a homogeneous system at $j/j_c=1.2$ as a function of frequency $\nu$. Two branches with decreasing magnitude of the amplitudeness measure are visible. Negative and positive values correspond to phase- and amplitude-like modes, respectively. The calculation was done in the Gutzwiller approximation (see Methods) for a square system with a length of $40$ sites applying periodic boundary conditions. \textbf{b,} Flatness $f=(a-b)/(a+b)$ of the trajectories of the order parameter in the complex plane as a function of frequency $\nu$. Again, two branches corresponding to phase-like (starting from $f<0$) and amplitude-like (starting from $f>0$) modes are visible. \textbf{c,} Illustration of the flatness parameter $f$ showing typical trajectories for the order parameter $\Psi$ in the complex plane. For both panels the mean value of $\Psi$ is $0.47$ along the real axis (small cross). For $f<0$ (left panel), the trajectory is an ellipse around the mean value, with its major axis perpendicular to the real axis (i.e., the direction of the order parameter). For $f>0$ (right panel), the major axis is parallel to the real axis.}}
\end{figure} 
The same generic behaviour of the amplitudeness measure is also seen in a homogeneous system (see Fig.\,12a). Two branches with negative and positive measure are visible, corresponding to phase-like and amplitude-like excitations. The amplitudeness measure is also related to the time evolution of the respective modes given by the trajectories of the order parameter $\Psi$ in the complex plane.  For a given mode, the trajectory is an ellipse around the equilibrium value of $\Psi$ (see Fig.\,12c) and the full ellipse is passed in a time $1/\nu$, where $\nu$ is the excitation frequency of the mode. To characterize the ellipses, we introduce the flatness parameter $f=(a-b)/(a+b)$, where $a$ and $b$ are the axes of the ellipse as defined in Fig.\,12c. The flatness parameter $f$ (Fig.\,12b) also shows two branches corresponding to phase- and amplitude-like modes, with a similar behaviour as the amplitudeness measure. In the respective branches, the modes with the lowest energies correspond to low-momentum modes. For these modes, the trajectories are very flat ellipses (large $|f|$) with their major axis perpendicular ($f<0$, phase-like) or parallel ($f>0$, amplitude-like) to the direction of the mean order parameter. With increasing momentum the modes become more circular, with a flatness parameter close to zero.

The main difference to the trapped system is the possibility to excite certain modes in the spectrum with lattice modulation. In an infinitely large homogeneous system, only modes with zero momentum can be excited with lattice modulation (the lowest energy mode of the respective branches). In a large enough trapped system, this selection rule is still approximately valid in the sense that only a few modes with the lowest energies of the respective branches couple significantly (see Fig.\,10b). The spatial pattern of these modes show only long-wavelength variations (equivalent to low-momentum in a trapped system). 
}
\end{document}